\documentstyle[12pt,osa]{revtex}
\begin{document}


\title{Homogeneous States of Finite Periodic CA with GKL Rules}         
\author{Ru-Fen Liu\footnote[2]{fmliu@phys.ncku.edu.tw} and Chia-Chu Chen\footnote[3]{chiachu@phys.ncku.edu.tw}}   
\date{National Cheng-Kung University\\ Physics department\\Tainan Taiwan ROC}               
\maketitle
\pagestyle{plain}
\begin{abstract}
\indent

For any finite lattice of size $2N$, it is shown that the GKL
automaton will evolve into attractors if the initial
configurations contains an homogeneous block of size larger than
$N$. The best time estimation for reaching these attracting states
is obtained. Our results also confirm  to the case of infinite
lattice which has been obtained previously.

\end{abstract}
\newpage
\section{Introduction}       
\indent \ \ \ \ $\bf Cellular$ $\bf Automata(CA)$ are simple
mathematical models which can generate complex dynamical
phenomena. In principle, $\bf CA$s are spatially-extended discrete
dynamical systems which consist of a lattice of sites. The state
of each site is in one of the $g$ states at time t and  has local
interactions. The system evolve synchronously in  discrete time
steps according to identical rules. Cellular Automata was
introduced by von Neumann in order to address the problems of
self-reproduction and evolution\cite{Neu}. However, in spite of
their mathematical interests, $\bf CA$ has not attracted much
attention until the introduction of John Conway's game of
lifes\cite{Con} around 1970. About twenty years ago, Stephen
Wolfram introduced the Cellular automata to the physics community
as models of complex dynamical systems \cite{Wol} and also as a
new approach to the parallel computing scheme(For a review see
Ref.$4$). The interest in $\bf CA$'s potential application
continues to grow. In fact Cellular Automata have been used to
simulate, for example, solar flares \cite{Isl}, fluid
dynamics\cite{Fri}, crystal growth\cite{Mac}, traffic
flow\cite{Fuk} and galaxy formation\cite{Ger}.
\newline
\indent One of the important question in $\bf CA$ studies is to
determine the behavior of the system at large times. According to
Wolfram's classification\cite{Wol} the asymptotic behavior of $\bf
CA$ can be classified into four classes which are : (1) spatially
homogeneous state, (2) fixed or periodic temporal structure and
(3) chaotic pattern throughout space. The fourth class of $\bf CA$
behaves in a much more complicated manner and was conjectured by
Wolfram as capable for performing universal computation.
\newline
\indent Due to the complexity of cellular automata, numerical
computations of time evolution become the main scheme in $\bf CA$
studies. In Wolfram's work\cite{Wol} some statistical approaches
was introduced to characterize quantitatively the patterns
generated by $\bf CA$ evolution. Statistical methods such as
entropies and dimensions can only provide average properties of
cellular automata. However, exact results are always needed in
\newpage
\noindent more elaborate discussions . To make this point more
explicit let us consider the effects of noise on $\bf CA$. Due to
the fact that all physical systems are coupled to noisy
environment,
\noindent it is interesting to know if the above classification of
the attractors of $\bf CA$ remains intact under the influence of
noise. But this question can only be answered by knowing the
deterministic $\bf CA$ exactly, and as a result exact evaluation
of the attractor without noise is called for. Unfortunately such
exact calculations only exists for a few cases. One of the most
studied case is the GKL automaton proposed in the 80'\cite{Gac}
and surprisingly the detail analytic proof was completed about a
decade later\cite{Sa}.
\newline
\indent The analytical proof of GKL automaton constructed in
Ref.11 was under the assumption that the system is infinite in
size. However all numerical investigation is studied on finite
lattice with periodic boundary condition. It is therefore
important to investigate GKL model with periodic lattice. In this
work we revisit this problem by modifying the method of Ref.11 and
prove that the GKL automaton defined on a finite periodic lattice
of size $2N$ can only have two uniform states if the initial
configurations containing an uniform block of size larger than
$N$. The best time estimation for reaching these uniform states is
obtained. Our results also include previous proof \cite{Sa} as a
special case.
\newline
\indent
 The plan for this paper is as follows. In the next section a
 brief review of the GKL automaton is given and several
 definitions are introduced. The main results of this work are given in terms
 of lemmas and theorem which are presented in section three.
A brief discussion of our results are provided at the last
section.

\section{ The GKL Automaton on Finite Periodic Lattice}       
\indent \ \ \ \ The cellular automaton is defined on a finite size
lattice with $2N$ sites. Any general configuration of CA will be
denoted by $\sigma$. The configuration of the system at time $t$
is denoted by $\sigma^t=\{\sigma^t(x)\mbox{, } 1 \leq x \leq 2N\}$
where $\sigma^t(x)$ which denotes the state on site
\newpage
\noindent $x$ takes on two possible values $\{+,-\}$. The periodic
boundary condition is realized by $\sigma^t(2N+k)$= $\sigma^t(k)$.
The GKL rules are given by the following updating procedures:
\\

\begin{eqnarray}\sigma^{t+1}(x)=\left\{\begin{array}{ccccc}majority\{\sigma^t(x),
\sigma^t(x+1), \sigma^t(x+3)\}&&if &\sigma^t(x)&
{=+}\\majority\{\sigma^t(x), \sigma^t(x-1), \sigma^t(x-3)\}&&if
&\sigma^t(x)&{=-}\end{array}\right.\end{eqnarray}
\newline
\indent  Any configuration $\sigma$ can always be divided into two
blocks which are denoted by $\sigma_B$ and $\sigma_I$. $\sigma_B$
is a block of size $N_B$ , which contains $\sigma(1)$ to
$\sigma(N_B)$, and $\sigma_I$ has size $N_I$, which contains
$\sigma(N_B+1)$ to $\sigma(2N)$. In this work, $N_B$ is required
to satisfy $N_B\geq N+1$ such that $N_B>N_I$. The time evolution
of $\sigma_B$ is denoted by $\sigma^t_B$. In this work a block
containing the same sign is called homogeneous.  Furthermore, the
initial configuration $\sigma^0$ always begins with a $\sigma_B$
which is a homogeneous contains only $+$ signs whereas $\sigma_I$
can have arbitrary configuration . We will show in the next
section that starting with this initial configuration the system
will end up in an attractor which is homogeneous for the whole
configuration with $\sigma^t=\{\sigma(x)=+\mbox{, } 1 \leq x \leq
2N\}$ in the limit of large $t$. That is to say that the block
$\sigma_I$ will be washed out at the end. Even though during time
evolution the process can be quite complicate. This particular
homogeneous configuration will be denoted by $\sigma_+$. A similar
attractor also exists for the case with initial condition
containing a $\sigma_B$ which is an homogeneous block of $-$ sign.
This attractor is denoted by $\sigma_-=\{\sigma(x)=-\mbox{, } 1
\leq x \leq 2N\}$. Before we proceed it is useful to introduce
some definitions which are in close resemblance to the ones
employed in Ref.$11$ but with modification.

\newtheorem{Def}{Definition}
\begin{Def}
\ \  $\sigma$ is good to $a$ if it satisfies three conditions as
follows:
\newline \indent(1) For  $1\leq x \leq a-2$, if $\sigma(x)=-$, then
$\sigma(x+1)=+$. \newline \indent (2) For  $1\leq x \leq a-4$, if
$\sigma(x)=-$, then $\sigma(x+3)=+$. \newline \indent(3) If
 $\sigma(a-3)=\sigma(a)=-$, then $\sigma(a-1)=-$.
\end{Def}
\newpage
\indent With this definition, it is clear that any initial
configuration, $\sigma^0$ is good to $N_B+2$. Furthermore, to the
left side of $a$, the configuration $\sigma$ satisfying the above
conditions contains neither two consecutive $-$ nor two $-$ at a
distance 3, except near the boundary site a. That is, $\{++--\}$
is allowed if the rightmost site of this block is $a$. For
example, the configuration $\sigma$ with

\begin{eqnarray} \sigma=\{-+-+++--++\cdots\} \nonumber\end{eqnarray}
is good to site $8$, but not good to any site larger than $8$. It
is convenient to denote \{$--$\} as $\bar{\sigma}\{x;2\}$ with
$\sigma(x)=\sigma(x+1)=-$. This notation can be generalized as
$\bar{\sigma}\{x;L\}$ which denotes a block containing only $-$ of
size $L$ where $L \geq 2$ and starting from $\sigma(x)$ to
$\sigma(x+L-1)$. Besides, we also denote \{$-++-$\} as
$\tilde{\sigma}\{x\}$ with $\sigma(x)=\sigma(x+3)=-$.

\begin{Def}
\ \ $\sigma$ is good if $\sigma$ is good for all sites.
\end{Def}
\begin{Def}
\ \ A block in $\sigma$ is perfectly good if this block contains
neither $\bar{\sigma}\{x;L\}$ nor $\tilde{\sigma}\{x\}$.
\end{Def}

\section{The Proof}       
\indent \ \ \ \ The main result in this section is to establish
the fact that the system has two attractors which are the
homogeneous states $\sigma_+$ and $\sigma_-$ respectively. First
we show that perfectly good configuration $\sigma$ stays as
perfectly good as the system evolves. This is state as a lemma:
\newtheorem{Lem}{Lemma}
\begin{Lem}
\ \ If $\sigma^t$ is perfectly good, then $\sigma^{t+1}$ is also
perfectly good.
\end{Lem}

\noindent \textbf{Proof:} It can be verified by direct checking
that if $\sigma^{t+1}$ is not perfectly good, then $\sigma^t$ is
also not perfectly good. For example, if $\sigma^{t+1}$ is not
perfectly good then it might \newpage \noindent contain
$\bar{\sigma}\{m;L\}$(A similar discussion can also be done for
$\tilde{\sigma}\{k\}$.). For this configuration if
$\sigma^t(m+1)=+$, then $\sigma^t(m)$ should be $-$ such that
either $\sigma^t(m-1)$ or $\sigma^t(m-3)$ is equal to $-$. As a
result, whatever the sign of $\sigma^t(m-2)$ be, $\sigma^t$ must
contain either $\bar{\sigma}\{m-1;2\}$ or $\tilde{\sigma}\{m-3\}$.
The same argument can also apply for the case of $\sigma^t(m+1)=-$
which implies the existence of either $\bar{\sigma}\{m-1;2\}$ or
$\tilde{\sigma}\{m-2\}$ in $\sigma^t$. Therefore, for
$\sigma^t(m+1)=\pm$, $\sigma^t$ is also not perfectly good.
\\\\
\indent It is useful to keep in mind that if $\sigma^t(m)=-$ and
$\sigma^{t+1}(m)$ is also $-$, then $\sigma^t$ must contains
either $\bar{\sigma}\{m-1;2\}$ or $\tilde{\sigma}\{m-3\}$ on the
\textit{\textbf{left}} side of $m$. It is also interesting to know
that the number of time steps for $\sigma_B$ to retain perfectly
good can be obtained and the result is proved in the following
lemma.

\begin{Lem}
\ \ For any initial configuration with an homogeneous $\sigma_B$,
the perfectly good configuration of $\sigma_B$ is retained during
time evolution within $N_B-2$ time steps.
\end{Lem}
\textbf{Proof:} It is noted that, if $\sigma^{t+1}_B$ contains
either $\bar{\sigma}\{m;L\}$ or $\tilde{\sigma}\{m\}$, then either
$\bar{\sigma}\{m';L'\}$ or $\tilde{\sigma}\{m'\}$ are contained in
$\sigma^t_B$. The proof of this fact can be established in the
same way as the proof of lemma 1 and is omitted.
\newline\indent However there is one particular case which requires
further discussion.  As $\sigma^{t+1}_B$ contains
$\bar{\sigma}\{3;L\}$ or $\tilde{\sigma}\{3\}$, there is a
possibility that $\sigma^t$ contains $\tilde{\sigma}\{2N\}$. If
that happens then the proof is invalidate. To show that this
particular case does not occur one has to establish that starting
from any homogeneous $\sigma_B$, $\sigma^t_B$ does not contain any
$\bar{\sigma}\{3;L\}$ or $\tilde{\sigma}\{3\}$ as $t\leq N_B-2$.
\newline \indent It is obvious that, the homogeneous configuration
$\sigma_B$ can only be eroded from the right side at most one site
at each time step. As a result, either $\bar{\sigma}\{3;L\}$ or
$\tilde{\sigma}\{3\}$ may appear at $t=N_B-2$. For the case of
$\bar{\sigma}\{3;L\}$, according to this at-most-one-site
propagating property, the state of any site less than $4$ must be
$+$ at $t=N_B-3$. Hence, $\sigma^{N_B-3}(3)=+$ with
$\sigma^{N_B-3}(4)=\sigma^{N_B-3}(6)=-$. However this can not lead
to $\bar{\sigma}\{3;L\}$ at next time step. As a result,
$\sigma^{N_B-2}$ does not contain $\bar{\sigma}\{3;L\}$. On the
other\newpage\noindent  hand, for the existence of
$\tilde{\sigma}\{3\}$, by using the propagating property, one can
conclude in a similar fashion that $\sigma^{N_B-3}(3)=+$ and
$\sigma^{N_B-3}(4)=\sigma^{N_B-3}(6)=-$ which imply
$\sigma^{N_B-3}(5)$ must be $-$. Such configuration is
contradicted to the definition of $\tilde{\sigma}\{3\}$. Therefore
$\sigma^{N_B-2}$ also does not contain $\tilde{\sigma}\{3\}$. This
completes the proof of this lemma.
\\\\ \indent
Combining the results of lemmas 1 and 2, one can conclude that
within $N_B-2$ time steps, if $\sigma_B$ at time $t$ is perfectly
good, then $\sigma_B$ at time $t+1$ is also perfectly good.
Moreover, Because our initial configuration of $\sigma_B$ is
perfectly good, then in $N_B-2$ time steps, $\sigma_B$ retains its
perfectly good configuration. We also mention that there is
another way to see why \textbf{lemma 2} only holds for $t\leq
N_B-2$. As the $-$ sign propagates to the site $3$ at $t=N_B-2$,
$\sigma^{N_B-1}(3)$ is determined by the site of
$\sigma^{N_B-2}(2N)$ which is contained in $\sigma_I$ and implies
$\sigma^{N_B-1}(2)=\sigma^{N_B-1}(3)=-$ could occur. If that is
the case then $\sigma_B$ is not perfectly good.
\begin{Lem}
\ \ In $N_B-2$ time steps, if $\sigma^t$ is good to $a$, then
$\sigma^{t+1}$ is good to $a+1$.
\end{Lem}
\textbf{Proof:} This lemma can be established by noting that if
$\sigma^{t+1}$ is not good to $a+1$, then $\sigma^t$ is not good
to $a$. For example, one can check that if any
$\bar{\sigma}\{m;L\}$ appears in $\sigma^{t+1}$, then
$\sigma^{t+1}$ is not good for all sites with $x\geq m+2$. By
using the same approach as one employed in proving the previous
lemmas it can be shown that $\sigma^t$ must contain either
$\bar{\sigma}\{m-1;2\}$ or $\tilde{\sigma}\{m-3\}$. This result
implies $\sigma^t$ is at least not good to $m+1$. This conclusion
can also be drawn for $\sigma^t(m+1)=-$. The case of
$\tilde{\sigma}\{k\}$ can also be treated in the same way and the
conclusion remains the same as in the case of
$\bar{\sigma}\{m;L\}$. According to our definition of perfectly
good, \textbf{lemma 2} ensures that $\sigma_B$ keeps its perfectly
good configuration in $N_B-2$ time-steps therefore $Lemma$ $3$ is
only true for $t\leq N_B-2$.

\begin{Lem}
\ \ If $\sigma^t$ is perfectly good and
$\sigma^{t}_{seed}\{m,L_s\}$ (with $L_s\geq 3$) is a homogeneous
block of $+$ sign in $\sigma^t$, then the length ($L_s$) of this
block increases at least two sites at the next time step.
\end{Lem} \newpage\noindent
\textbf{Proof:} It can be proved directly. When $\sigma^t$ is
perfectly good, then it can contain any block which has all $+$
signs with length other than $2$. It is a trivial case for the
block containing only one single $+$ thus we just need to consider
any block with the length larger than two. If $\sigma^t$ contains
a block $\sigma^t_{seed}\{m,L_s\}$, then $\sigma^{t+1}$ contains a
homogeneous block of $+$ sign starting from $\sigma^{t+1}(m)$ to
$\sigma^{t+1}(m+L_s-2)$. Now consider the case with
$\sigma^{t}(m-1)=-$, due to the restriction of perfectly good of
$\sigma^t$, it is easy to check that $\sigma^{t}(m-2)$ and
$\sigma^{t}(m-4)$ must be $+$. Hence
$\sigma^{t+1}(m-1)=\sigma^{t+1}(m-2)=+$. Furthermore, from
\textbf{lemma 1}, $\sigma^{t+1}$ is also perfectly good therefore
if $\sigma^{t}(m-3)=-$, it must change sign with
$\sigma^{t+1}(m-3)=+$ at the next time step. However, the reader
can easily check that there is no constraint on
$\sigma^{t+1}(m-4)$ and $\sigma^{t+1}(m+L_s-1)$. As a result,
$\sigma^t_{seed}\{m;L_s\}$ becomes
$\sigma^{t+1}_{seed}\{m-3,L_s+2\}$ in $\sigma^{t+1}$.
\\\\\indent
Equipping with these lemmas, one can now proceed to prove that the
system will evolve into the attractor if the initial configuration
has an homogeneous $\sigma_B$ with $N_B=N+1$.
\\\\
\textbf{Theorem}\ \ \textit{For $\sigma$ with a size of $2N$, if
the initial configuration has homogeneous $\sigma_B$ with $N_B\geq
N+1$ then $\sigma^{N+N_I-2}=\sigma_+$.}
\\\\
\textbf{Proof:} For our initial configuration of $\sigma_B$,
$\sigma^0$ is good to $N_B+2$. According to \textbf{lemma 3},
$\sigma^{N_I-2}$ is good to $2N$. It is easy to check that the
rightmost four sites of $\sigma^{N_I-2}$ of all good
configurations have $+$ sign. These sites are on the left side of
$\sigma_B$ (since the lattice is periodic.). Hence, in the next
time step, one obtains
$\sigma^{N_I-1}(2N-1)=\sigma^{N_I-1}(2N-2)=+$. Furthermore,
according to \textbf{lemma 2}, $\sigma_B$ still maintains its
perfectly good configuration at time $N_I-1$ and therefore
$\sigma_I$ contains neither $\bar{\sigma}\{m;L\}$ nor
$\tilde{\sigma}\{m\}$. Thus, $\sigma^{N_I-1}$ is perfectly good.
Moreover, at this time step, $\sigma_B$ also contains a block
$\sigma_+$ with at least of size $N_B-(N_I-1)\geq 3$. As a result,
there is at least a block $\sigma^{N_I-1}_{seed}\{1,3\}$ in
$\sigma^{N_I-1}$ with a perfectly good configuration.
\newpage\noindent From \textbf{lemma 4},
$\sigma^{N_I-1}_{seed}\{1,3\}$ will expand with the expansion rate
of $2$ sites per each time step. Therefore, the number of time
step for $\sigma^{N_I-1}$ to become $\sigma_+$ is at most
$\frac{2N-3+1}{2}$. As a result, with our particular initial
configuration of a finite size lattice $\sigma$,
$\sigma^{N_I+N-2}=\sigma_+$. Thus the time for reaching the
homogeneous state is obtained and is given by $T\leq N_I+N-2$.
\section{Discussions and Conclusions}
\indent \ \ \ \
     The results obtained in this work is that for any finite one
dimensional lattice with periodic condition, the GKL automaton has
only two homogeneous attractors if the initial configuration
contained a homogeneous block of size bigger than half of the
lattice. These results are closely related to the previous results
which were obtained for infinite lattice\cite{Sa}. The previous
result studied the eroding effects which wash out any finite
island embedded in any infinite homogeneous configuration. It is
shown in this work that the infinite homogeneous configuration is
not necessary. In fact the infinite homogeneous configuration can
be replaced by a finite homogeneous block where elimination of the
island also occurs as long as the homogeneous block is larger than
half of the lattice. Furthermore the upper bound of the time for
reaching the attractor is also obtained in this work. Even though
the results obtained here is an improvement of the earlier
results, however the story is not complete yet. This is due the
fact that, on finite lattice, numerical simulations\cite{Mit} has
shown that for any initial configuration with a density of either
$+$ or $-$ sign greater than $0.5$ the system always ends up on
the $\sigma_\pm$ attractors respectively. The results in this work
certainly confirm to this fact. Therefore it will be interesting
to rigorously establish the results of simulation by relaxing the
condition of a homogeneous $\sigma_B$. Work along this line is now
being investigated.

\newpage

\

\end{document}